# High-Specific-Power Flexible Transition Metal Dichalcogenide Solar Cells

Koosha Nassiri Nazif,[1,†] Alwin Daus,[1,†] Jiho Hong,[2,3] Nayeun Lee,[2,3] Sam Vaziri,[1] Aravindh Kumar,[1] Frederick Nitta,[1] Michelle Chen,[3] Siavash Kananian,[1] Raisul Islam,[1] Kwan-Ho Kim,[4,5] Jin-Hong Park,[4,6] Ada Poon,[1] Mark L. Brongersma,[2,3,7] Eric Pop,[1,3] and Krishna C. Saraswat[1,3*]

[1]Dept. of Electrical Engineering, Stanford University, Stanford, CA 94305, USA

[2]Geballe Laboratory for Advanced Materials, Stanford University, Stanford, CA 94305, USA

[3]Dept. of Materials Science and Engineering, Stanford University, Stanford, CA 94305, USA

[4]Dept. of Electrical and Computer Engineering, Sungkyunkwan University, Suwon 16419, Korea

[5]Dept. of Electrical and Systems Engineering, University of Pennsylvania, Philadelphia, PA 19104, USA

[6]SKKU Advanced Inst. of Nanotechnology (SAINT), Sungkyunkwan University, Suwon 16419, Korea

[7]Dept. of Applied Physics, Stanford University, Stanford, CA 94305, USA

[†]These authors contributed equally.

*corresponding author email: saraswat@stanford.edu

**Abstract:** Semiconducting transition metal dichalcogenides (TMDs) are promising for flexible high-specific-power photovoltaics due to their ultrahigh optical absorption coefficients, desirable band gaps and self-passivated surfaces. However, challenges such as Fermi-level pinning at the metal contact-TMD interface and the inapplicability of traditional doping schemes have prevented most TMD solar cells from exceeding 2% power conversion efficiency (PCE). In addition, fabrication on flexible substrates tends to contaminate or damage TMD interfaces, further reducing performance. Here, we address these fundamental issues by employing: 1) transparent graphene contacts to mitigate Fermi-level pinning, 2) $MoO_x$ capping for doping, passivation and anti-reflection, and 3) a clean, non-damaging direct transfer method to realize devices on lightweight flexible polyimide substrates. These lead to record PCE of 5.1% and record specific power of 4.4 W $g^{-1}$ for flexible TMD ($WSe_2$) solar cells, the latter on par with prevailing thin-film solar technologies cadmium telluride, copper indium gallium selenide, amorphous silicon and III-Vs. We further project that TMD solar cells could achieve specific power up to 46 W $g^{-1}$, creating unprecedented opportunities in a broad range of industries from aerospace to wearable and implantable electronics.

Conventional silicon (Si) solar cells dominate the photovoltaics market with a market share of about 95% due to their low-cost manufacturing and reasonable power conversion efficiency (PCE).[1] However, the low optical absorption coefficient and brittle nature of Si lead to degraded performance in ultrathin, flexible Si solar cells and therefore prevent their broader usage in applications demanding high power per weight (i.e.

specific power, $P_S$) and flexibility, for example in aerospace, transportation, architecture and self-powered wearable and implantable electronics.[2–10]

Emerging semiconducting transition metal dichalcogenides (TMDs) exhibit excellent properties for such flexible high-specific-power photovoltaics. These include ultrahigh optical absorption coefficients up to one order of magnitude greater than conventional direct band gap semiconductors, near-ideal band gaps for solar energy harvesting, and self-passivated surfaces.[11–18] In fact, ultrathin (<20 nm) TMDs can achieve near-unity, broadband, and omnidirectional absorption in the visible spectrum.[15,16] The wide range of TMD band gaps (~1.0 eV to 2.5 eV)[17] are also well suited for highly efficient single-junction or double-junction tandem solar cells.[13] In addition, the dangling-bond-free surfaces of layered TMDs enable heterostructures without the constraint of lattice matching, offering abundant design choices for TMD photovoltaics.[18] According to realistic detailed balance models,[13] a PCE of ~27% can be achieved in ultrathin single-junction TMD solar cells, leading to extremely high $P_S$ once implemented on lightweight flexible substrates.[8]

Despite these promising forecasts, there have not been any such demonstrations due to difficulties in reaching high PCE and integrating materials on flexible substrates. The PCE of TMD solar cells has typically not exceeded 2%,[19–25] mostly due to strong Fermi-level pinning at the metal contact-TMD interface[26] and the inapplicability of traditional doping schemes such as diffusion or ion implantation, which can damage TMDs.[27] Reducing or eliminating Fermi-level pinning by adopting a gentle metal transfer method,[22,26] introducing an ultrathin interlayer at the metal-TMD interface,[28–31] or forming a van der Waals (vdW) heterojunction such as graphene-TMD[32,33] can significantly improve the performance of TMD devices. In addition, forming a p-n homojunction by employing TMD-compatible doping methods such as surface charge transfer and fixed charge doping via metal oxides,[23–25] plasma doping,[34] or electrostatic doping[35] has resulted in remarkable performance improvements. Noteworthily, the highest PCEs in thin-film single-junction TMDs are 2.8% in plasma-doped $MoS_2$ and 6.3% in electrostatically-doped $MoSe_2$ solar cells.[34,35] At the same time, TMDs are typically transferred to flexible substrates and most of these processes can damage TMD interfaces, leave unwanted polymer residues, and do not allow for a reliable and practical vertical device architecture.[36] Previous reports on $P_S$ of TMD solar cells, i.e. 3 W g$^{-1}$ with a PCE of 0.46%[22] and 2500 W g$^{-1}$ with a PCE of 1.0%,[14] do not account for the substrate's weight, which practically constitutes the largest part of the overall weight and needs be considered for accurate $P_S$ calculations. The only TMD solar cell on a lightweight, flexible substrate reported to date has a PCE of <0.7%, yielding a $P_S$ of <0.04 W g$^{-1}$.[20]

Here, we address the above-mentioned device and integration challenges by utilizing transparent graphene contacts mitigating Fermi-level pinning,[32,33] $MoO_x$ capping for doping, passivation and anti-reflection coating,[24,37,38] and a clean, non-damaging direct transfer method to realize devices on an ultrathin (5 μm), lightweight and flexible polyimide (PI) substrate.[39] The flexible TMD (WSe$_2$) solar cells made in this fashion achieve a PCE of 5.1%, surpassing previous flexible TMD solar cells by more than an order of magnitude.[20] Furthermore, the integration on an ultrathin substrate enables a $P_S$ of 4.4 W g$^{-1}$, more than 100x higher than previous results on flexible TMD photovoltaics[20] and in the same range as champion solar cells of prevailing thin-film technologies cadmium telluride (CdTe), copper indium gallium selenide (CIGS), amorphous silicon (a-Si) and group III-V semiconductors.[40–47] In future, TMD solar cells on even thinner substrates and with higher PCEs could potentially achieve an unprecedented $P_S$ of ~46 W g$^{-1}$ (as we project in this work) opening up far-reaching possibilities in a broad range of industries.[9]

We fabricate flexible vertical photovoltaic cells from multilayer (~200 nm) tungsten diselenide (WSe$_2$) absorbers, transparent hole-collecting graphene top contacts covered by MoO$_x$ doping, passivation and anti-reflection coatings, and optically-reflective electron-collecting gold (Au) bottom contacts. The bottom contact and absorber material are embedded into a flexible, transparent PI substrate. Device schematics and optical images are shown in Fig. 1a-d. We mechanically exfoliate WSe$_2$ flakes on thermally oxidized silicon substrates and deposit patterned Au bottom contacts, which are all covered with spin-coated PI and released together in deionized (DI) water.[39] The patterned transparent top contacts constituted of graphene and MoO$_x$ are then formed on top via graphene wet transfer and MoO$_x$ electron-beam (e-beam) evaporation. Details on device fabrication and transfer procedures can be found in Supplementary Information Section S1.

Figure 1e shows the schematic energy band diagram of flexible WSe$_2$ solar cells based on values for WSe$_2$, Gr, and Au reported in the literature. WSe$_2$ (bulk band gap of ~1.2 eV and electron affinity of ~4.0 eV)[21,48–50] is considered "undoped", with a low hole concentration of ~10$^{14}$ cm$^{-3}$ due to the presence of Se vacancies (information provided by the bulk crystal vendor). Due to the energetic nature of e-beam evaporation, defect states are induced at the Au–WSe$_2$ interface and the Au Fermi level is pinned toward the charge neutrality level of WSe$_2$ located at midgap.[26,51,52] This decreases the effective work function of Au and makes it a decent electron-collecting contact. We find that replacing Au with lower work function metals such as Ti and Al leads to a lower performance, most probably due to their reactive nature therefore forming poor interfaces with WSe$_2$ (see Supplementary Information Section S2).[53] On the other hand, layered materials Gr and WSe$_2$ experience no Fermi-level pinning at their vdW interface.[32,33] The work function of undoped graphene is ~4.6 eV (e.g. in vacuum), which increases to ~5.0 eV when graphene is exposed to air,[37,54,55] forming a Schottky junction with the "undoped" WSe$_2$. The MoO$_x$ on top of graphene further increases its work function by doping it p-type,[37] therefore enhancing the built-in potential of the Gr–WSe$_2$ Schottky junction. MoO$_x$ also passivates the top surface of the solar cell.[24] These lead to a higher open-circuit voltage ($V_{OC}$) and short-circuit current density ($J_{SC}$) in MoO$_x$-capped WSe$_2$ solar cells (Supplementary Information Section S3). As we will discuss later in the optical characterization section,

MoO$_x$ also serves as an effective anti-reflection coating for WSe$_2$, leading to an additional increase in $J_{SC}$. Given the approximate location of Gr, WSe$_2$ and Au Fermi levels, the depletion regions of Gr–WSe$_2$ and Au–WSe$_2$ Schottky junctions are estimated to be in the order of 1 μm and therefore expand throughout the entire depth of the ~200-nm-thick WSe$_2$ layer, leading to fully depleted devices.

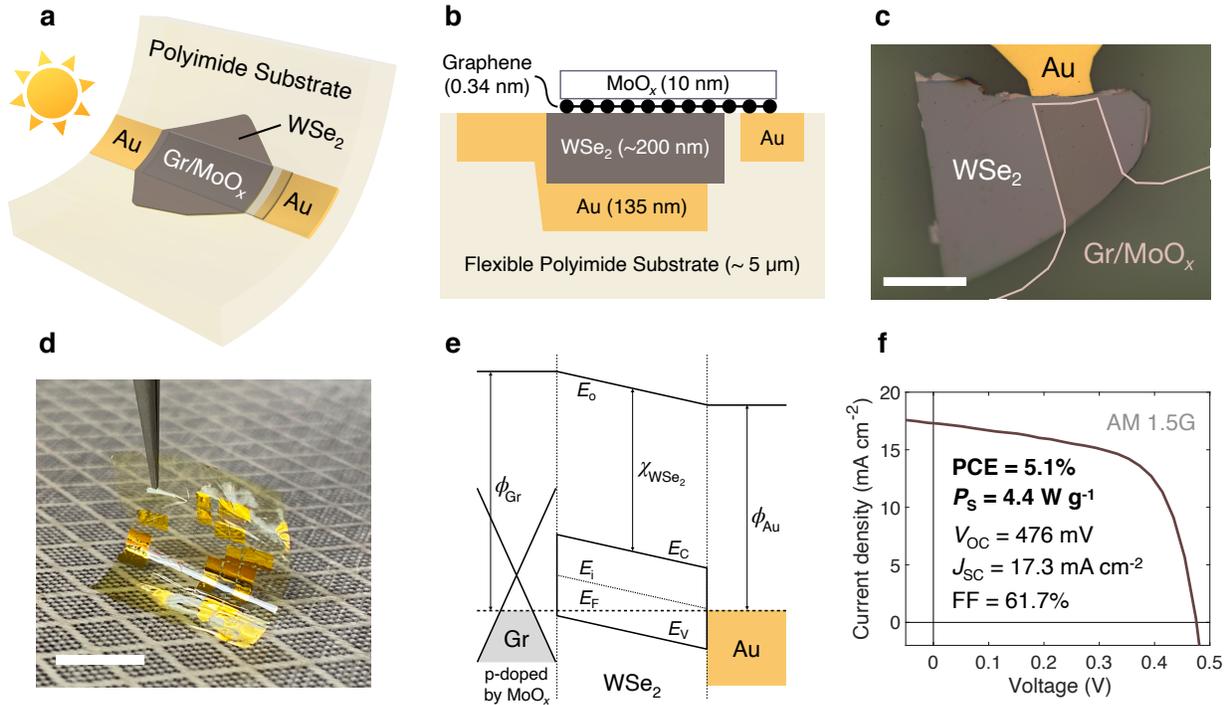

**Figure 1 | Flexible WSe$_2$ solar cells. a,** Device schematic and **b,** cross-section. **c,** Top-down optical image of the device. Scale bar, 50 μm. **d,** Photograph of WSe$_2$ solar cells on a flexible polyimide substrate. Scale bar, 1 cm. **e,** Qualitative energy band diagram of the device. The Fermi level ($E_F$) is pinned near midgap at the Au-WSe$_2$ interface, but not at the van der Waals graphene (Gr)–WSe$_2$ interface.[26,32,33,51,52] MoO$_x$ increases the Gr work function and the built-in potential of the Gr–WSe$_2$ Schottky junction.[37] $E_0$, vacuum level; $E_C$, conduction band edge; $E_V$, valence band edge; $E_i$, intrinsic Fermi level; $\Phi_{Au}$, Au effective work function; $\Phi_{Gr}$, Gr work function; $\chi_{WSe_2}$, electron affinity of WSe$_2$. **f,** Measured current density $J$ vs. voltage $V$ under AM 1.5G illumination. PCE, power conversion efficiency. $P_S$, specific power or power per weight. $V_{OC}$, open-circuit voltage. $J_{SC}$, short-circuit current. FF, fill factor.

Under global air mass AM 1.5G illumination, the flexible WSe$_2$ solar cells achieve $V_{OC}$ of 476 mV, a $J_{SC}$ of 17.3 mA cm$^{-2}$ and a fill factor (FF) of 61.7% (Fig. 1f), leading to an unprecedented PCE of 5.1% in flexible TMD solar cells, over 10x higher than previous demonstrations (<0.7%).[20] Having an ultra-thin absorber layer and a lightweight PI substrate, these WSe$_2$ solar cells also achieve a high specific power ($P_S$) of 4.4 W g$^{-1}$ (calculated in Supplementary Information Section S4), over 100x higher than preceding results (<0.04 W g$^{-1}$)[20] and on par with champion solar cells from well-established thin-film technologies CdTe,

CIGS, a-Si and III-Vs.[40–47] The device has a shunt resistance of 226 Ω cm$^2$ and a series resistance of 3.1 Ω cm$^2$, calculated by inversing the slope of the *J*−*V* curve at short-circuit and open-circuit conditions, respectively, yielding the reasonable FF of 61.7%. We measure reproducible performance in all nine devices fabricated (Supplementary Fig. S5). No hysteresis is observed in the *J*–*V* characteristics when sweeping the voltage in the forward and backward directions (Supplementary Fig. S6). Similar *J*–*V* characteristics are observed in solar cells with "undoped" tungsten disulfide ($WS_2$) absorber layers. However, due to the higher work function of $WS_2$ compared to $WSe_2$ and therefore the lower built-in potential in Gr–$WS_2$ Schottky junctions, $WS_2$ solar cells exhibit lower $V_{OC}$, $J_{SC}$, FF and hence PCE (see Supplementary Fig. S7).

Next, we measure the current density vs. voltage (*J*–*V*) characteristics of flexible $WSe_2$ solar cells in the dark and for AM 1.5G illumination at various incident power intensities (Fig. 2a–b). As shown in Fig. 1e, Au is not an ohmic n-type contact and, similar to Gr, forms a Schottky junction with $WSe_2$, resulting in a back-to-back diode structure (Fig. 2a, inset). This undesirable Schottky barrier at the Au back contact leads to the roll-over phenomenon where the slope of *J*–*V* curve is reduced at high forward biases,[56,57] starting here at around *V* = 0.65 V. The barrier also causes the cross-over of dark and light *J*–*V* curves at *V* = 0.53 V, which can be explained by the presence of a minority carrier surface recombination current at the Au–$WSe_2$ interface.[57] The two phenomena are also frequently observed in CdTe and CIGS solar cells.[56–58]

Fig. 2b shows a zoomed-in view of the photovoltaic region. An analysis of this data indicates that the shunt resistance decreases almost linearly with increasing incident intensity (see Supplementary Fig. S8). This phenomenon, known as photoshunting, occurs due to increased minority carrier conductivity across the device under illumination.[59,60] Improving the charge carrier selectivity of the solar cell, for example by utilizing carrier-selective metal-interlayer-semiconductor (MIS) contacts or introducing a high built-in potential p-n homojunction could reduce or eliminate photoshunting. Given the initially high shunt resistance of the device, photoshunting does not affect the shape of the *J*–*V* curve and therefore fill factor stays constant at various intensities.

By fitting a power law equation on the measured current density and incident power data (Fig. 2c), we observe that short-circuit current density versus incident power follow a linear trend ($J_{SC} = \beta \cdot (P_{in})^\alpha$, $\alpha = 1$), expected from a well-designed solar cell. Equation 1 is a rearrangement of the diode equation in the presence of photogeneration ($J_{photo} = J_{SC}$) and shunt resistance ($R_{SH}$) at $V = V_{OC}$, leading to zero current density by definition (*J* = 0). In this equation, *n* is the diode ideality factor, $k_B$ is the Boltzmann constant, *T* is the absolute temperature, *q* is the elementary charge, and $J_o$ is the dark saturation current. According to this equation, the open-circuit voltage scales linearly with $\ln(J_{SC} - V_{OC}/R_{SH})$ when $(J_{SC} - V_{OC}/R_{SH})/J_o \gg 1$, valid for the WSe$_2$ solar cells in this study. By fitting the measured $V_{OC}$ and $J_{SC} - V_{OC}/R_{SH}$ (Fig. 2d), we extract $n$ and $J_o$ of the WSe$_2$ solar cells, neglecting the Au back Schottky diode for simplicity.

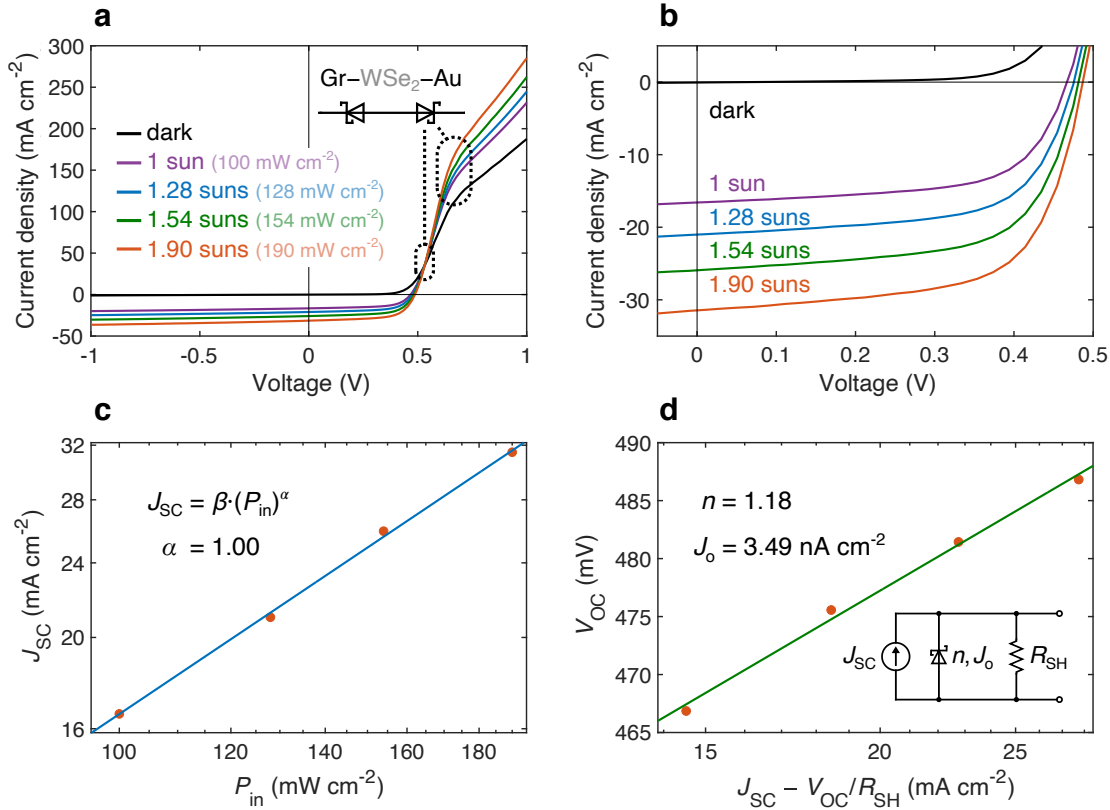

**Figure 2 | Electrical characteristics of flexible WSe$_2$ solar cells. a,** Current density vs. voltage (*J*–*V*) characteristics of WSe$_2$ solar cells under AM 1.5G illumination, at various incident power. Inset represents the circuit diagram of Au–WSe$_2$ and Gr–WSe$_2$ junctions. With its Fermi level pinned at midgap, Au forms a Schottky junction to WSe$_2$ opposing the main Gr–WSe$_2$ Schottky junction, leading to cross-over and roll-over effects occurring around 0.53 V and 0.65 V, respectively,[56,57] as marked by dotted ovals. **b,** Zoomed-in view of the photovoltaic region of the *J*–*V* measurements shown in part (a). **c,** Short-circuit current density (*J*$_{SC}$) of the devices. *P*$_{in}$, incident power; symbols, measurements; line, power law fit. **d,** Open-circuit voltage (*V*$_{OC}$) of the devices. *R*$_{SH}$, shunt resistance; symbols, measurements; line, fit. Inset shows a representative circuit diagram. *n* is the ideality factor and *J*$_o$ the dark saturation current density from the diode fit.

WSe$_2$ solar cells demonstrate a desirable near-unity ideality factor of $n = 1.18$ and dark saturation current of $J_o = 3.49$ nA cm$^{-2}$. The near-unity ideality factor and small dark saturation current indicate low levels of charge carrier recombination and therefore a high internal quantum efficiency as confirmed later by comparing the measured $J_{SC}$ and $J_{SC,\,max}$ derived from absorption measurements.

$$V_{OC} = \frac{nk_BT}{q}\ln\left(\frac{J_{SC} - \frac{V_{OC}}{R_{SH}}}{J_o} + 1\right) \tag{1}$$

Fig. 3a shows the optical image, spatial light beam induced current (LBIC or photocurrent) map acquired at a wavelength of 530 nm, and their overlay for a typical flexible WSe$_2$ solar cell. The overlay map shows that only the Gr–WSe$_2$ diode is responsible for splitting photogenerated holes and electrons and therefore producing photocurrent. No photocurrent generation is observed at the Au–WSe$_2$ back diode. This can be seen at the narrow WSe$_2$ region near the Au contact line on the left, where the Au bottom contact is present, but no Gr is covering the WSe$_2$. In contrast, a strong photocurrent is measured on the opposite side of the WSe$_2$ on the right, where the Au back contact is absent and WSe$_2$ is only in contact with Gr. This is further visualized by a cross-shaped contact scheme in Supplementary Fig. S9.

To accurately define the active area of the device, the photocurrent profile across the width of the device (x-axis in Fig. 3a) is plotted on a linear scale (Fig. 3b). In this specific device, MoO$_x$ is slightly misaligned with respect to Gr (see Fig. 3a). The misalignment only occurred in few devices due to lithography issues. Most devices, such as the one shown in Fig. 1f, have well aligned Gr and MoO$_x$. On the left edge of Fig. 3b (Gr, corresponds to the upper edge in the photocurrent map), photocurrent is only generated in regions covered by Gr. The tail beyond the Gr edge occurs due to the finite laser spot size (~2 μm), leading to spatial averaging of the photocurrent. This spatial averaging shows up as a ~2-μm tail going from non-zero to zero photocurrent regimes as visible in the photocurrent profile. On the right side of Fig. 3b (corresponds to the lower edge in the photocurrent map), due to the passivation effect of MoO$_x$,[24] current generation goes beyond Gr and occurs up to the MoO$_x$ right edge. A similar spatial averaging phenomenon is also taking place on this side, resulting in a ~2-μm tail beyond the MoO$_x$ edge. The photocurrent profile confirms that photogeneration only occurs in regions covered by Gr (and MoO$_x$, if misaligned) and this area can be used to accurately define the active area of the device for current density calculation, similar to other studies on vertical TMD solar cells.[22] The active area of the solar cells tested vary from ~10$^3$ to ~10$^4$ μm$^2$.

We measure the absorption spectrum of WSe$_2$ solar cells at different stages of fabrication, i.e. after polyimide release (Au–WSe$_2$), after Gr transfer (Au–WSe$_2$–Gr) and finally after MoO$_x$ deposition (Au–WSe$_2$–Gr–MoO$_x$), as shown in Fig. 3c. For consistency, each measurement is taken at exactly the same spot at the center of the active area of the device. The data in Fig. 3c corresponds to the device whose *J–V* characteristics are shown in Fig. 1f. This device has a 209-nm-thick WSe$_2$ absorber layer, as measured by a stylus-based surface profiler.

After transferring Gr on top of WSe$_2$, the overall absorption of the stack is slightly reduced. Optical simulations using the transfer matrix method produce a similar result (Supplementary Fig. S10a). Depositing 10 nm of MoO$_x$ on top of Gr increases the overall absorption of the stack. This can be either due to parasitic absorption within MoO$_x$ or its anti-reflection coating effect improving the absorption within the WSe$_2$ absorber layer. To answer this question, we simulate absorption using the transfer matrix method.

Fig. 3d shows simulated absorption spectrum of the Au–WSe$_2$–Gr–MoO$_x$ stack along with the contribution of each individual layer. Simulated and measured absorption spectra are in good agreement, having the same shapes and magnitudes, with peaks and valleys located at similar wavelengths. The small discrepancies between the two absorption spectra can be explained by the fact that the optical properties of WSe$_2$ used in the simulation are taken from the literature[61] and may deviate slightly from the WSe$_2$ films in this study.

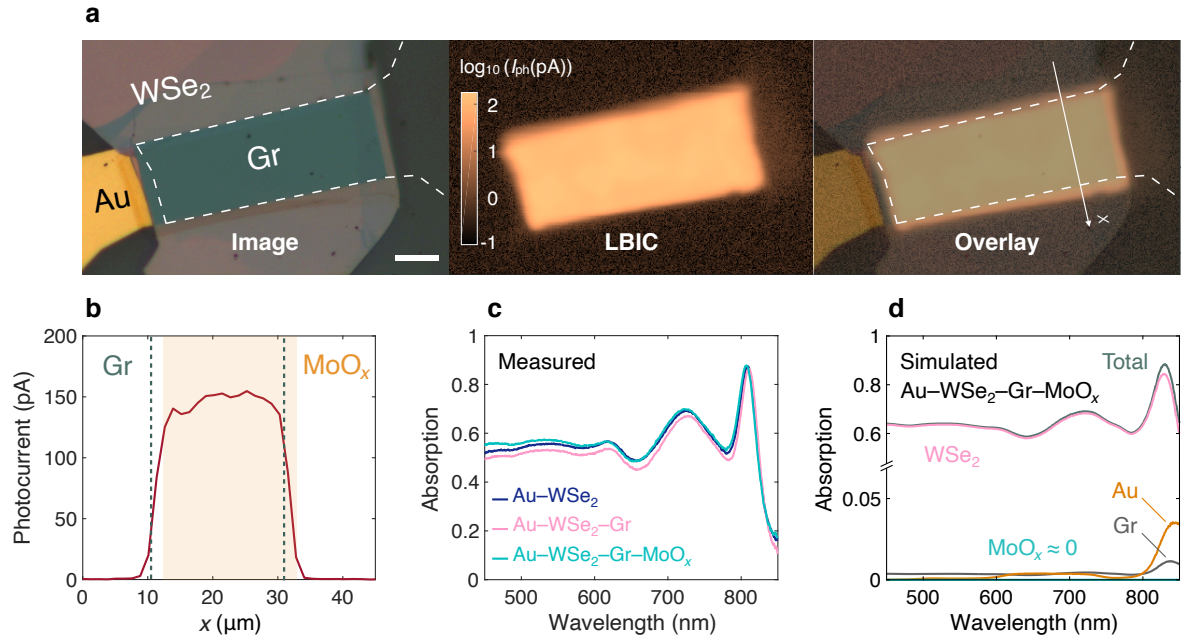

**Figure 3 | Optical characteristics of flexible WSe$_2$ solar cells. a,** Optical image (left), light beam induced current (LBIC or photocurrent) map (middle), and their overlay (right) for a typical WSe$_2$ solar cell, measured at λ = 530 nm. Scale bar, 10 µm. **b,** Photocurrent profile along the x-axis shown in part (a), demonstrating that the Gr/MoO$_x$–WSe$_2$ junction area accurately represents the active area of the solar cell. The misaligned MoO$_x$ is passivating the WSe$_2$ top surface,[24] leading to current generation up to the edge of MoO$_x$. The photocurrent tails beyond the edges of Gr and MoO$_x$ occur due to the finite laser spot size (~2 µm). In most devices MoO$_x$ and Gr are fully aligned like in Fig. 1c. **c,** Measured absorption spectra at the center of the active area, taken at various stages of the fabrication process: after polyimide release (only with Au–WSe$_2$), Gr transfer and MoO$_x$ deposition. **d,** Simulated absorption spectra of the same device, matching well with the measurement in part (c). The plot also shows the contribution of each layer to the overall absorption. Nearly all absorption occurs within the 209-nm-thick WSe$_2$ absorber layer. Note the discontinuous vertical axis, used to magnify the smaller contributions.

According to Fig. 3d, parasitic absorption in the 10-nm MoO$_x$ layer is essentially zero. In addition, nearly all absorption occurs within the 209-nm-thick WSe$_2$ absorber layer, indicating that the absorption boost observed after MoO$_x$ coating (Fig. 3c) is mainly due to the increased absorption in the WSe$_2$ layer. These observations suggest that MoO$_x$ is acting as an anti-reflection coating for the WSe$_2$ absorber layer. Our optical simulation confirms this hypothesis (Supplementary Fig. S10b-c), showing that MoO$_x$ increases the absorption within the WSe$_2$ layer. The simulation also reveals that an optimal choice of MoO$_x$ thickness (~70 nm) can lead to a significant improvement in WSe$_2$ absorption, resulting in $J_{SC}$ values up to 30 mA cm$^{-2}$ (Supplementary Fig. S10c-d). This suggests that MoO$_x$ could be used as a simple yet effective anti-reflection coating choice for TMD photovoltaics, to be further investigated in future studies.

The WSe$_2$ solar cells show an average optical absorption of about 55% over the 450–850 nm wavelength spectrum. Using the simulated WSe$_2$ absorption spectrum and assuming unity internal quantum efficiency (IQE), we calculate a maximum $J_{SC}$ of 20.0 mA cm$^{-2}$, slightly underestimated because absorption at wavelengths below 400 nm and above 1000 nm are excluded due to lack of available material data (see Supplementary Fig. S10d). Given the measured J$_{SC}$ of 17.3 mA cm$^{-2}$ (Fig. 1f), this implies an average IQE (weighted by AM 1.5G spectrum) of 0.87, which signals low levels of charge carrier recombination, in agreement with near-unity ideality factor and small dark saturation current extracted from $J$–$V$ measurements (Fig. 2d). Similar IQE values have been observed in other vertical Schottky junction TMD solar cells.[22]

To test the performance of devices under bending, we attach the PI substrate onto an 8-mm-diameter metal cylinder, which bends the substrate at a curvature radius of 4 mm (Fig. 4a). The flexible WSe$_2$ solar cells show the same $J$–$V$ characteristics in flat and bent states under AM 1.5G illumination (Fig. 4b), indicating consistent performance levels under bending. This is not surprising because given the PI substrate thickness of only 5 μm, the materials encounter small strain values of ~0.06% at this bending radius.[39] We have investigated similar TMD, metal and dielectric stacks in electronic devices in more detail in our recent work and found that there is no discernable change of electrical device properties.[39] For our solar cells with small exfoliated flakes on length scales that are tens of microns, there is no effect of substrate curvature on the light-coupling, as seen in Fig. 4b. In future, if the active area of the solar cell is increased e.g., by large-area synthesis of TMDs, these bending studies will become more important to quantify the effects of the bent surface on light-coupling and thus $J_{SC}$, which would alter solar cell performance.

Fig. 5 benchmarks the PCE and $P_S$ of the flexible WSe$_2$ solar cells in this work against other thin-film solar technologies (details in Supplementary Section S11). The flexible WSe$_2$ solar cells in this study demonstrate remarkable improvements of about 10x and 100x in PCE and $P_S$, respectively, compared to the previous results in flexible TMD solar cells (PCE <0.7% and $P_S$ <0.04 W g$^{-1}$).[20] With only a moderate PCE of 5.1%, the WSe$_2$ solar cells already achieve a high specific power of 4.4 W g$^{-1}$ enabled by their ultrathin WSe$_2$ absorber layer and lightweight PI substrate. This high $P_S$ is in the same range as champion solar cells of well-established thin-film technologies CdTe (II-VI), CIGS, a-Si, and III-Vs.[40–47]

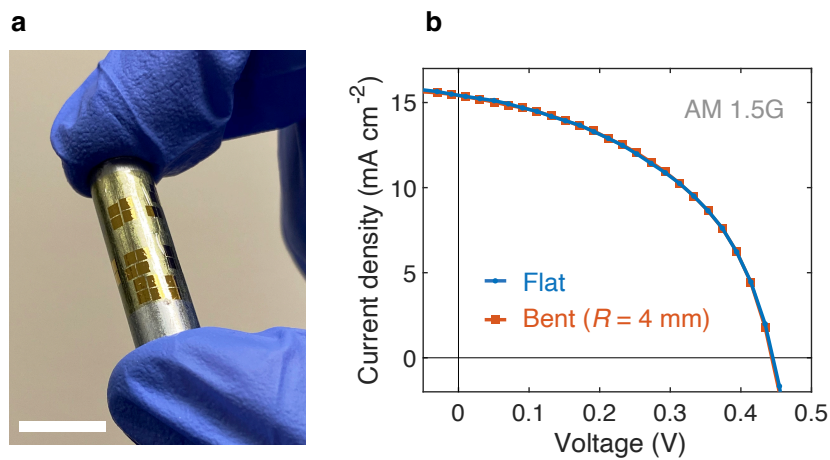

**Figure 4 | Bending test. a,** Photograph of the bending setup. The polyimide substrate is attached to an 8-mm-diameter metal cylinder, causing bending of the substrate at a curvature radius of 4 mm. Scale bar, 1 cm **b,** Measured *J–V* characteristics of a typical flexible WSe$_2$ solar cell under AM 1.5G illumination in flat and bent conditions. Bending does not change the *J–V* characteristics of the device.

By reducing the PI substrate thickness to 1 μm, similar as in some of the champion organic PV (OPV) and perovskite solar cells in Fig. 5, specific power can be further increased to 8.6 W g$^{-1}$ (path #1 in Fig. 5). According to realistic detailed balance models developed for TMD photovoltaic cells, single-junction multilayer TMDs can in principle achieve ~27% PCE with an optimized optical and electronic design.[13] Such PCE would lead to an ultrahigh specific power of 46 W g$^{-1}$ (path #2 in Fig. 5), by far outperforming all other thin-film technologies, including perovskite solar cells which currently hold the record for specific power (29.4 W g$^{-1}$).[62] In addition, TMD solar cells do not have the stability challenges of OPV or perovskites, and in contrast to high-performing perovskites and lead sulfide (PbS) quantum dots, they do not contain toxic elements such as lead, and therefore are not expected to pose any significant environmental or health hazards.[63]

In order to achieve the projected ~27% PCE[13] in the flexible WSe$_2$ solar cells both optical and electronic designs need to be improved. As pointed out earlier, MoO$_x$ can serve as an effective anti-reflection coating for WSe$_2$. Our optical simulation shows that simply increasing the thickness of MoO$_x$ to an optimal value of ~70 nm is supposed to improve the absorption within the WSe$_2$ layer to 80% and enable $J_{SC}$ values up to 30 mA cm$^{-2}$ (see Supplementary Fig. S10c-d). Metasurface-based plasmonic light trapping schemes can help further improve absorption and reach $J_{SC}$ values near the Shockley-Queisser limit (40 mA cm$^{-2}$ for a band gap of 1.2 eV).[64–68]

$V_{OC}$ is another important area of improvement. The built-in potential and therefore $V_{OC}$ of these devices can be improved by employing n-type WSe$_2$. The doping process can be performed during growth, or by means of metal oxides such as AlO$_x$ and TiO$_x$.[69,70] Replacing Au with a lower work function metal but ensuring a similar interface quality could also improve $V_{OC}$. We showed in our experiments that Al and Ti are not good candidates for this purpose (see Supplementary Information Section S2). Forming a high built-in potential p-n homojunction in the TMD absorber layer is another way to achieve a high $V_{OC}$, possibly by p-type (MoO$_x$) and n-type (AlO$_x$) doping on the top and the bottom.[23–25,38,69,71] One can also improve the $V_{OC}$ by adopting carrier selective (MIS) contacts which both de-pin the Fermi level and enable a selective collection of only one type of charge carrier on each side of the solar cell.[28–31,72,73]

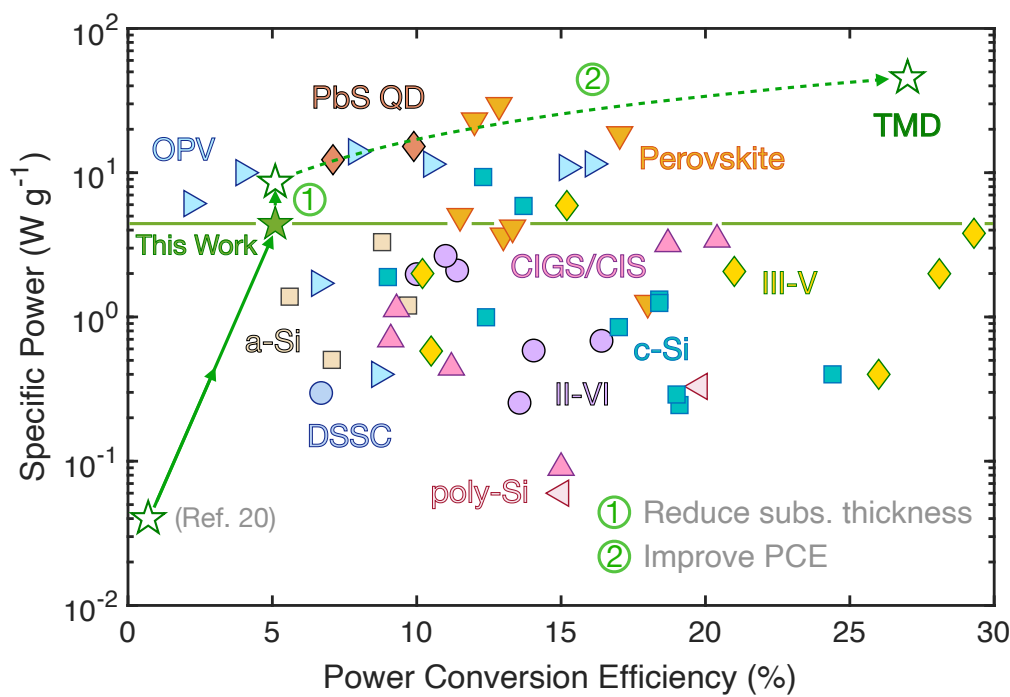

**Figure 5 | Power conversion efficiency (PCE) and specific power (power per weight) of lightweight and flexible thin-film solar technologies.** Our flexible TMD (WSe$_2$) solar cell achieves a relatively high specific power despite its moderate PCE (filled green star). Arrow 1 shows projected effect of reducing substrate thickness, arrow 2 shows projected effect of improving PCE. With these improvements, TMD solar cells could reach unprecedented specific power in the future. More details are given in Supplementary Information S11. OPV, organic photovoltaics; PbS QD, lead sulfide quantum dot; CIGS/CIS, copper indium (gallium) selenide; c-Si, crystalline silicon; a-Si, amorphous silicon; poly-Si, polycrystalline silicon; DSSC, dye-sensitized solar cell; III-V (II-VI), compound semiconductors containing elements from groups three and five (or two and six) in the periodic table (See Supplementary Table S3 for more details and references).

Research efforts to scale up TMD growth to large areas would soon enable scalable and low-cost production of TMD photovoltaic cells,[74,75] similar to other chalcogenide solar cells CdTe and CIGS. The potential of TMDs to achieve high power conversion efficiency and specific power at a low cost as well as their stability and environmentally-friendliness (in contrast to perovskites) makes them a serious candidate for next-generation photovoltaics, especially in high-specific-power applications.

In summary, we demonstrated flexible WSe$_2$ solar cells with record-breaking power conversion efficiency PCE of 5.1% and power per weight $P_S$ of 4.4 W g$^{-1}$. We performed detailed optical and electrical characterizations on these solar cells to explain their superior performance and identify areas of improvement, providing practical guidelines on the optical and electronic design to enhance PCE and $P_S$. We also tested the flexible solar cells under bending and showed similar performance in flat and bent states. Future large-area TMD cells will require more bending studies to verify effects of light-coupling at different angles on the solar cell performance. Lastly, we benchmarked the flexible WSe$_2$ solar cells against other thin-film photovoltaic technologies and showed their potential to achieve ultrahigh $P_S$, creating unprecedented opportunities in a broad range of industries from aerospace to wearable electronics.

**Methods**

**Device fabrication**: The details on device fabrication are provided in Supplementary Section S1. In short, WSe$_2$ flakes were mechanically exfoliated onto Si/SiO$_2$ substrates. Next, 135 nm of Au serving as solar cell bottom contacts was electron-beam evaporated and structured by optical photolithography and lift-off. Then the flexible PI substrate was spin-coated on top, cured and all structures were released together in DI water. After flipping the substrate, graphene was wet-transferred as transparent top contact and structured by optical photolithography and dry etching. Finally, MoO$_x$ was deposited by electron-beam evaporation and structured by optical photolithography and lift-off, followed by annealing in ambient air for the purpose of doping, passivation and anti-reflection coating. The process flow is schematically shown in Supplementary Fig. S1. Thickness of WSe$_2$ flakes was measured by atomic force microscopy (AFM, Bruker Dimension Icon) and/or stylus-based surface profilometry (Alphastep 500) after exfoliation on the Si/SiO$_2$ substrate.

**AM 1.5G current–voltage measurements**: AM 1.5G *I–V* measurements were done using a digital source meter (Keithley 2420) and a class AAA solar simulator (Newport, Oriel Sol3A Class AAA) having a 450 W xenon short arc lamp and AM 1.5G spectral correction filter. Lamp intensity calibration was done using a silicon reference cell (Newport, Oriel 91150V) placed at the location of the sample. The silicon reference cell was calibrated by Newport Corporation. *I−V* characteristics were measured with a scan rate of 200 mV s$^{-1}$ and a dwell time of 30 ms. The measurements were performed in air. The samples were kept at room temperature via convection cooling provided by a fan.

**Photocurrent mapping**: The photocurrent from the device was measured on a custom-built optoelectronic setup. A supercontinuum laser source (Fianium) and an acousto-optic tunable filter (Fianium) were used to provide monochromatic illumination across a broad spectral range. To achieve high signal-to-noise ratio, the laser light was modulated by a chopper wheel (400 Hz) synchronized with two lock-in amplifiers. The laser light was focused onto the sample using a 50X long working distance objective (Mitutoyo M Plan APO NIR). To image the sample, two beam splitters were placed in the illumination path for a halogen lamp and a charge-coupled device (CCD) imaging camera, respectively. Using a glass slide, a small fraction of the reflected light was directed into a large-area Si photodiode (New Focus, model 2031) connected to a lock-in amplifier (Stanford Research SR810 DSP) to measure reflection from the sample. The sample was placed on a chip carrier and then mounted on a three-axis piezo stage to accurately control its spatial position. Electrical connections were made by soldering wires (34-gauge enameled wires) using thermally stable solder paste Sn42/Bi57.6/Ag0.4 T4 (Chip Quik) onto the Au metal pads. To measure the photocurrent, the mounted sample was connected in series with a source meter (Keithley 2612), a tunable current-to-voltage amplifier and a second lock-in amplifier (Stanford Research SR810 DSP).

**Absorption measurement**: The absorption measurements were performed using a Nikon C2 confocal microscope coupled to a CCD camera (Acton Pixis 1024, Princeton Instrument) and a spectrometer (Acton SP2300i, Princeton Instrument). Unpolarized light from a halogen lamp was used to illuminate the sample through a 20X objective (Nikon CFI Achromat LWD, NA = 0.4). The reflection spectra ($R(\lambda)$) were normalized to the calibrated reflection spectrum of a protected silver mirror (Thorlabs, PF10-03-P01). Since all devices were fabricated on top of thick Au contact, the absorption spectra were calculated by absorption($\lambda$) = 1 – reflection($\lambda$).

**Raman Spectroscopy**: The Raman measurements were performed on a HORIBA Scientific LabRAM HR Evolution spectrometer using an excitation wavelength of 532 nm. For Raman measurements an acquisition time, accumulations, laser power and optical grating of 20 s, 2, 2.8 mW, 1800 gr mm$^{-1}$ were used, and the spot size is less than 1 μm.

**Optical Simulation**: Optical simulations were performed using the transfer matrix method. A normally incident plane wave was assumed in all simulations. The thickness of each layer used in simulations is shown in Fig 1b. The bottom Au contact was modeled as a semi-infinite substrate due to its small penetration depth. Optical constants for $WSe_2$, Gr and Au were taken from the literature.[61,76,77] Optical constants for $MoO_x$ were obtained experimentally from spectroscopic ellipsometry (n ≈ 2.0).

**Authors Contributions:** K.N. and A.D. contributed equally. K.N., A.D. and S.V. conceived the project. K.N. and A.D. fabricated the devices, assisted by S.V., A.K. and M.C. J.H. carried out the photocurrent measurements, assisted by S.K. for electrical connections made via soldering. N.L. did the absorption measurements, assisted by J.H. K.N., A.D., and A.K. performed the *J–V*, Raman/ellipsometry, and AFM measurements, respectively. A.D. and K.N. did the surface profilometry measurements and mechanical bending tests. N.L. and J.H. carried out the optical simulations. F.N. and K.N. performed data analysis on *J–V* measurements. All authors contributed to the data interpretation, presentation, and writing of the manuscript. K.C.S. supervised the work.

**Acknowledgements:** Part of this work was performed at the Stanford Nanofabrication Facility (SNF) and Stanford Nano Shared Facilities (SNSF), supported by the National Science Foundation under award ECCS-2026822. The authors would like to thank SNF staff members for their help with the fabrication process. A.D. was supported by the Swiss National Science Foundation's Early Postdoc.Mobility fellowship (grant P2EZP2_181619), the Beijing Institute of Collaborative Innovation (BICI) and by the National Science Foundation (NSF) Engineering Research Center for Power Optimization of Electro-Thermal Systems (POETS) with Cooperative Agreement No. EEC-1449548. N.L., J.H and M.L.B acknowledge support from the Department of Energy Grant DE-FG07- ER46426. The authors acknowledge partial support from the member companies of the SystemX Alliance at Stanford.

**Competing Interests:** The authors declare no competing financial interests.

# Supplementary Information

## High-Specific-Power Flexible Transition Metal Dichalcogenide Solar Cells

Koosha Nassiri Nazif,[1,†] Alwin Daus,[1,†] Jiho Hong,[2,3] Nayeun Lee,[2,3] Sam Vaziri,[1] Aravindh Kumar,[1] Frederick Nitta,[1] Michelle Chen,[3] Siavash Kananian,[1] Raisul Islam,[1] Kwan-Ho Kim,[4,5] Jin-Hong Park,[4,6] Ada Poon,[1] Mark L. Brongersma,[2,3,7] Eric Pop,[1,3] and Krishna C. Saraswat[1,3,*]

[1]Dept. of Electrical Engineering, Stanford University, Stanford, CA 94305, USA
[2]Geballe Laboratory for Advanced Materials, Stanford University, Stanford, CA 94305, USA
[3]Dept. of Materials Science and Engineering, Stanford University, Stanford, CA 94305, USA
[4]Dept. of Electrical and Computer Engineering, Sungkyunkwan University, Suwon 16419, Korea
[5]Dept. of Electrical and Systems Engineering, University of Pennsylvania, Philadelphia, PA 19104, USA
[6]SKKU Advanced Inst. of Nanotechnology (SAINT), Sungkyunkwan University, Suwon 16419, Korea
[7]Dept. of Applied Physics, Stanford University, Stanford, CA 94305, USA
[†]These authors contributed equally.
*corresponding author email: saraswat@stanford.edu

This file includes:

- Section S1. Detailed fabrication process including transfer procedure
- Section S2. Back metal contact study
- Section S3. Doping effects of $MoO_x$
- Section S4. Specific power calculation
- Section S5. Reproducibility
- Section S6. Forward and backward scans
- Section S7. Flexible $WS_2$ solar cells
- Section S8. Photoshunting
- Section S9. Supplementary photocurrent mapping
- Section S10. Effects of Gr and $MoO_x$ on optical absorption
- Section S11. Benchmarking

**Section S1. Detailed fabrication process including transfer procedure**

We follow a recently developed fabrication and transfer approach,[39] where we perform the initial fabrication steps on a rigid Si/SiO$_2$ substrate followed by the release of patterned electrodes and the transition metal dichalcogenide (TMD) embedded into an ultrathin (~5 μm) flexible polyimide (PI) substrate (Fig. S1). This technique is advantageous for vertical device architectures as it eliminates large steps in surface topography leading to a flat surface for the subsequent transfer of transparent graphene top electrodes.

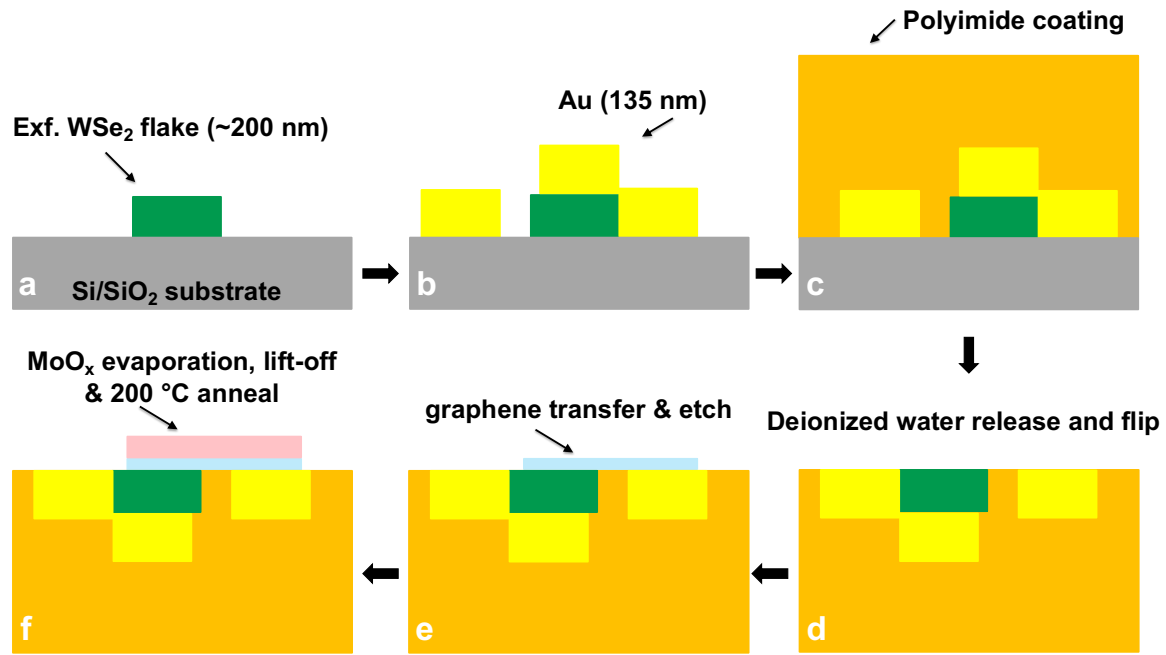

**Figure S1 | Flexible solar cell fabrication process**. **a,** Exfoliated WSe$_2$ flake on Si/SiO$_2$ substrate. **b,** Deposition of Au bottom contact and contact pads. **c,** Polyimide spin-coating. **d,** Free-standing polyimide substrate (flipped) after release in deionized water as in Ref. 39. **e,** Graphene transfer and etching. **f,** MoO$_x$ evaporation, patterning and anneal.

The detailed process steps are described below and the summarized process sequence is schematically shown in Fig. S1. First, 90 nm of SiO$_2$ was grown on a bulk silicon wafer by dry thermal oxidation at 1100 °C. The wafers were then manually cut into ~2 cm × 2 cm pieces. Intrinsically doped ("undoped" per manufacturer's description) WSe$_2$ flakes were mechanically exfoliated from the bulk crystal (2D Semiconductors) onto low-residue thermal release tape (Nitto Denko REVALPHA). They were subsequently transferred from the tape onto the Si/SiO$_2$ substrate using a WF film (Gel-Film®, WF-20-X4). Next, we spin-coated a lift-off layer LOL 2000 (3000 rpm, 60 s) on our substrates and baked it on a hot plate at 200 °C for 5 minutes. This was followed by spin-coating of Shipley 3612 photoresist (5500 rpm, 30 s) and baking on a hot plate at 90 °C for 1 minute. The lift-off layer/photoresist stack was then patterned by photolithography on a direct write lithography tool (Heidelberg MLA 150, dose: 90 mJ cm$^{-2}$, defocus: -2) and developed (MF-26A, 32 s). 135 nm of Au was deposited on top using electron-beam evaporation (AJA International). After a lift-off process similar as in Ref. 39, patterned Au bottom contacts and metal pads were left behind on top of the flakes and the substrate. The use of Au without any additional sticking layer enables us to later pick up the metal structures and TMD because no covalent bonds with the silicon dioxide substrate are formed (more details in Ref. 39).

The PI (PI-2610, HD MicroSystems) was spin-coated on top of the electrodes and TMD conformally covering all structures, baked at 90 °C and 150 °C for each 90 s, and finally cured at 250 °C in nitrogen ambient for 30 minutes. After that, all structures embedded into PI can be easily released from the silicon substrate by agitation and gentle mechanical force with a tweezer in deionized (DI) water without any discernable damage (more details in Ref. 39). After flipping the substrate, we temporarily attach it again to a silicon carrier with poly(methyl methacrylate) (PMMA) to perform the graphene transfer.

Graphene is grown via chemical vapor deposition at 1060 °C in an AIXTRON Black Magic furnace on commercial Cu foils with large preferentially 100 oriented grains (JX Mining, 99.9% purity HA-V2 treated rolled copper foil), which are surface cleaned in glacial acetic acid and commercial thin film nickel etchant (Transene, Nickel Etchant TFB) solution prior to growth.[78] After growth, the graphene is covered with PMMA via spin-coating (495-A2, 2500 rpm, 60 s, bake 140 °C for 45 s; 950-A4, 2000 rpm, 60 s, bake 80 °C 5 min). The backside of the Cu substrate is reactive ion etched in $O_2$ plasma (Oxford 80 RIE, 20 W, 10 mTorr, 40 sccm, 60 s). Then, the Cu/graphene/PMMA stack is dropped on top of $FeCl_3$ floating on the surface to etch the Cu from the backside. After that the graphene supported by PMMA is transferred by a glass slide into DI water (2 times) followed by diluted HCl:$H_2O_2$ solution (20:1:1 DI water:37% HCl:30% $H_2O_2$) and a final DI water beaker staying in each bath for 10 min. Then the graphene/PMMA layer is picked up by the supported PI substrate and the whole stack is allowed to gradually heat to 130 °C on a hotplate for drying. Finally, all PMMA (on top of graphene and below PI) is dissolved in toluene (2 hours) followed by acetone and iso-propanol soaks for each 5 min. To perform optical lithography on the free-standing PI substrates, we attached them to silicon carrier pieces prior to spin-coating with a few drops of photoresist (MEGAPOSIT™ SPR220-3) followed by nitrogen blow drying to flatten the substrate and remove excess photoresist and solvent, which was then baked on a hotplate at 90 °C for 5 minutes. The same photoresist was used for the optical lithography (spin-coating 3000 rpm, 30 s, bake 115 °C 90 s, exposure dose 250 mJ cm$^{-2}$ and defocus -2, post-exposure bake 115 °C for 90 s, develop in MF-26A 75 s). We then etched graphene to remove it in undesired areas by reactive ion etching with the same parameters as mentioned above.

For the purpose of doping, passivation and anti-reflection coating, 10 nm of patterned MoO$_x$ was deposited on top by e-beam evaporation (Kurt J. Lesker) from a MoO$_3$ pellet (Advanced Chemicals, 99.95% purity) at a rate of 0.3 Å s$^{-1}$, using the same photolithography and lift-off processes as described for the graphene etch and Au contact definition, respectively. Afterwards, we annealed the samples on a hot plate at 200 °C in ambient air for 10 minutes in order to further oxidize MoO$_x$ and therefore increase its work function, resulting in improved surface charge-transfer doping.[79]

**Section S2. Back metal contact study**

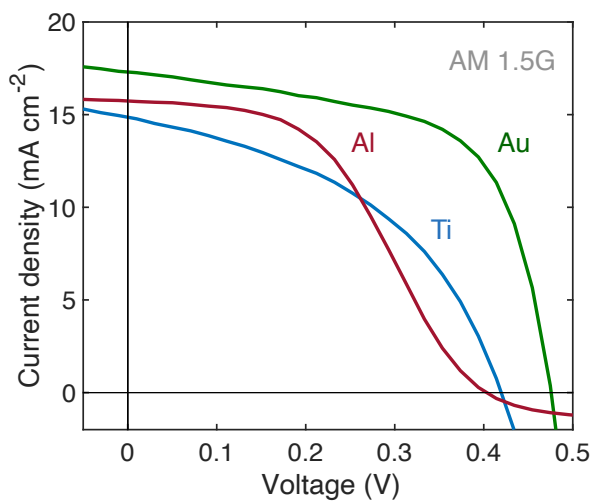

**Figure S2 | Performance of flexible WSe$_2$ solar cells having different back contact metals.** Typical $J$–$V$ characteristics of flexible WSe$_2$ solar cells having Al, Ti or Au back contact, under AM 1.5G illumination. Ti and Al, despite their lower work functions compared to Au, lead to worse performance, most probably due to their reactive nature and therefore forming poor interfaces with WSe$_2$. Ti and Al are known to be highly reactive/oxidizing, especially when in contact with TMDs, leading to the formation of AlO$_x$ at the Al–WSe$_2$ interface and TiO$_x$ + Ti$_x$Se$_y$ at the Ti–WSe$_2$ interface.[53] These defective interfaces result in high levels of charge carrier recombination and therefore decreased open-circuit voltage and short-circuit current. Insulating AlO$_x$ also induces high series resistance as well as a strong roll-over effect in Al-contact solar cells.

**Section S3. Doping effects of MoO$_x$**

The doping/passivation effect of MoO$_x$ on graphene and WSe$_2$ was analyzed by Raman spectroscopy (Table S1 and Fig. S3a-b). We find small peak shifts towards higher wavenumber for graphene after MoO$_x$ deposition, which has been previously reported to be a result of p-type doping and an increase of graphene's work function.[37] The WSe$_2$ peaks have much smaller yet discernable shift towards lower wavenumbers, also associated with material doping as shown in.[38] Overall, the two doping effects result in an increase in the built-in potential of the Gr–WSe$_2$ junction (more p-type doping in Gr than in WSe$_2$), leading to increased open circuit voltage ($V_{OC}$) and short-circuit current ($J_{SC}$). The passivation[24] and anti-reflection coating (Fig. S10b-d) effects of MoO$_x$ further increase the $J_{SC}$, leading to significant PCE improvements (Fig. S3c).

**Table S1 | Raman peak shifts upon deposition of MoO$_x$.** A total of 12 spots on several WSe$_2$ flakes were measured. Average Raman peak positions (cm$^{-1}$) for WSe$_2$ and graphene before and after coating with MoO$_x$ are compared in this table showing small peak shifts. The error bar represents the standard deviation.

| Peak | Before MoO$_x$ | After MoO$_x$ |
|---|---|---|
| **Graphene: G** | 1594.0 ± 3.9 | 1597.7 ± 5.9 |
| **Graphene: 2D** | 2689.2 ± 5.2 | 2702.0 ± 3.5 |
| **WSe$_2$: E/A** | 247.0 ± 0.2 | 246.7 ± 0.5 |
| **WSe$_2$: 2LA(M)** | 257.4 ± 0.3 | 257.2 ± 0.3 |

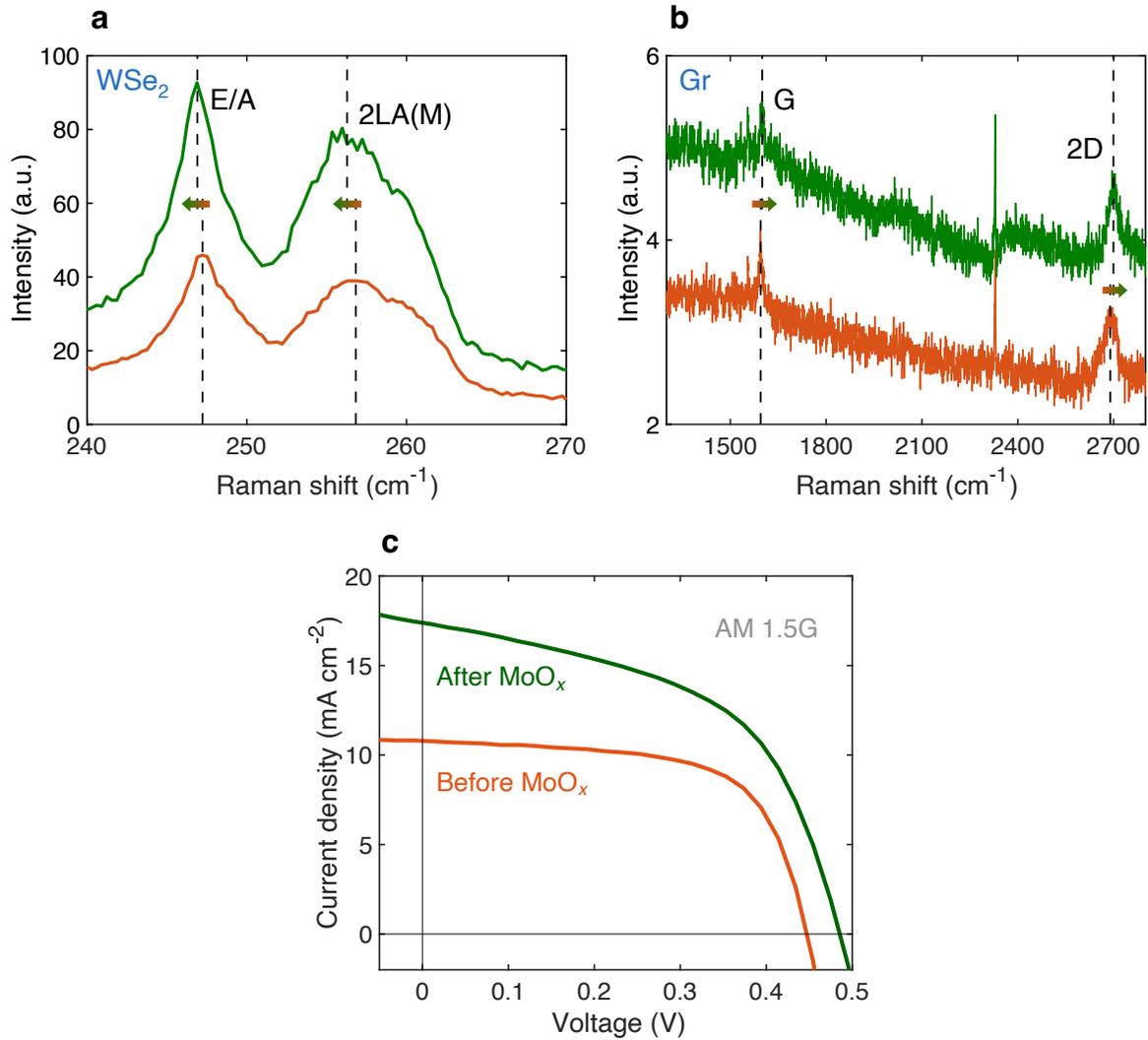

**Figure S3 | Performance-improvements by MoO$_x$.** Raman spectroscopy of **a,** WSe$_2$ and **b,** Gr in the Gr–WSe$_2$–Au–PI stack before and after MoO$_x$ deposition. Note, the narrow peak at ~2330 cm$^{-1}$ can be associated with N$_2$ vibrations from ambient air.[80,81] **c,** J–V characteristics of an example Au–WSe$_2$–Gr vertical solar cell before and after MoO$_x$ capping. PCE is significantly improved after MoO$_x$ capping due to doping, passivation and anti-reflection effects of MoO$_x$.

**Section S4. Specific power calculation**

The specific power ($P_S$) was calculated based on the solar cell efficiency and the areal weight densities of all materials in the solar cell stack including the PI substrate. Incident power is $P_{in}$ = 100 mW cm$^{-2}$, which corresponds to one-sun illumination. We can then calculate the maximum output power $P_{max}$ (unit: mW cm$^{-2}$) as follows:

$$P_{max} = J_{SC} \cdot V_{OC} \cdot FF = PCE \cdot P_{in}$$

where $J_{SC}$, $V_{OC}$, FF and PCE are the short-circuit current density, open-circuit voltage, fill factor and power conversion efficiency, respectively. Then, we sum up the areal mass densities of all materials in the solar cell stack including the substrate by using the volumetric mass density multiplied by the respective material thickness.

**Table S2 | Areal mass density of our solar cells**. The volumetric densities of all materials were taken from literature. For MoO$_x$ the volumetric mass density of MoO$_3$ was assumed.

| Peak | PI | Au | WSe$_2$ | Gr | MoO$_x$ | TOTAL |
|---|---|---|---|---|---|---|
| **Thickness (nm)** | 5000 | 135 | 200 | 0.3 | 10 | |
| **Volumetric mass density (g m$^{-3}$)** | 1.4×10$^6$ [82] | 1.93×10$^7$ [83] | 9.276×10$^6$ [84] | 2.267×10$^6$ [85] | 4.69×10$^6$ [86] | |
| **Areal mass density (g m$^{-2}$)** | 7 | 2.6055 | 1.864 | 0.00068 | 0.0469 | 11.517 |

We calculate $P_S$ (unit: W g$^{-1}$) by dividing $P_{max}$ by the total area mass density from Table S2. With $P_{max}$ = 5.1 mW cm$^{-2}$ (see Fig. 1f) we obtain $P_S$ = 4.4 W g$^{-1}$. A simple approach to reduce $P_S$ would be thinning down the substrate to ~1 μm, which is the approximate substrate thickness in some of the works with highest $P_S$ achieved so far,[62,87,88] leading to a value of ~8.6 W g$^{-1}$. A PCE of ~27% can be practically achieved in an optimized TMD single-junction as shown in,[13] further increasing the $P_S$ to ~46 W g$^{-1}$. We included these projections in Fig. 5 to emphasize the great potential of TMDs for high-specific-power photovoltaics.

**Section S5. Reproducibility**

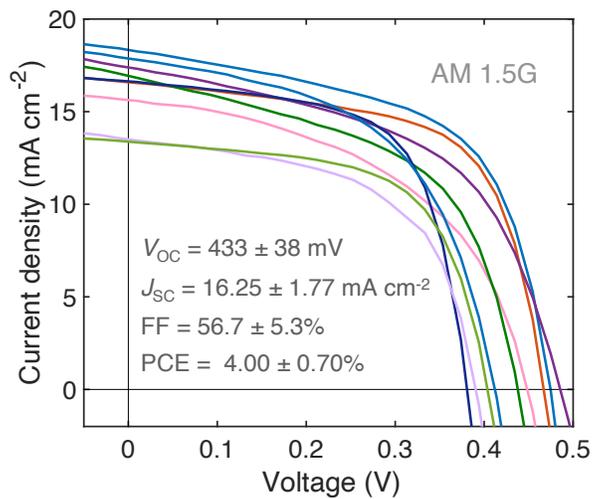

**Figure S5 | Reproducibility**. *J–V* measurements of 9 different flexible WSe$_2$ solar cells under AM 1.5G illumination, showing similar characteristics.

**Section S6. Forward and backward scans**

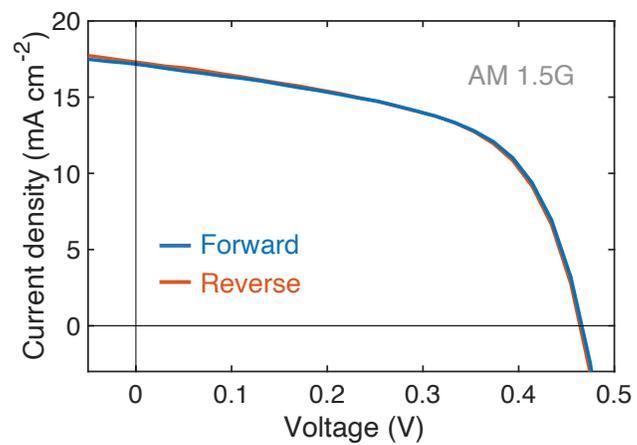

**Figure S6 | Forward and backward scans**. J-V characteristics under one-sun illumination show no hysteresis in forward/reverse voltage sweeps.

**Section S7. Flexible WS$_2$ solar cells**

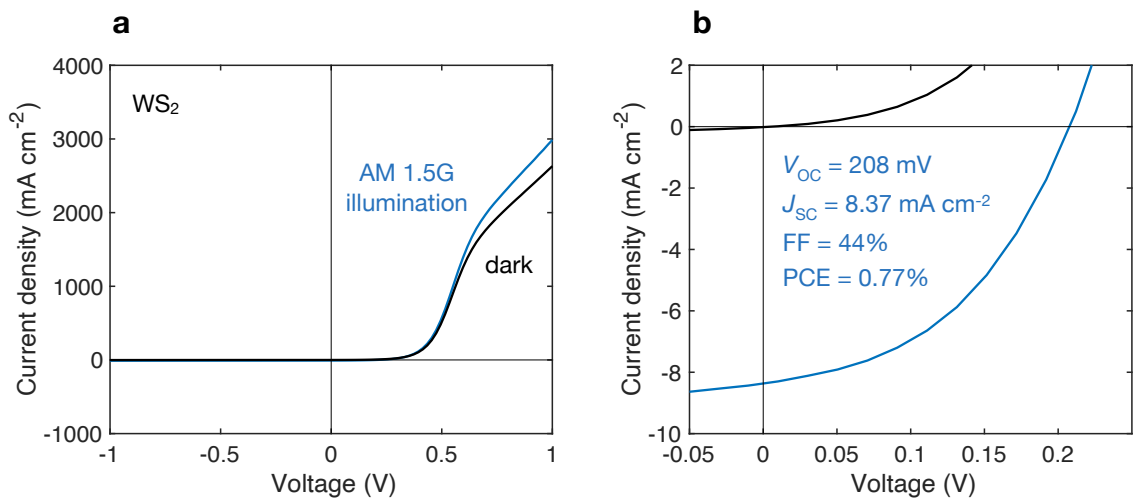

**Figure S7 | Flexible WS$_2$ solar cells**. **a-b,** *J–V* characteristics of flexible WS$_2$ solar cells having the exact same design as WSe$_2$ cells discussed in the main text (Fig. 1b), in dark and under one-sun illumination. **b,** a zoom-in view of the photovoltaic region. WS$_2$ solar cells show the same *J–V* characteristics as WSe$_2$ cells, however with lower performance. This is due to the higher work function of undoped WS$_2$ compared to undoped WSe$_2$ (higher electron affinity and bandgap in WS$_2$),[89] leading to a smaller built-in potential in WS$_2$–Gr Schottky junction and therefore reduced $V_{OC}$, $J_{SC}$, and FF. Further doping Gr (for example by using flame-deposited MoO$_3$)[37] is one way to achieve a similar built-in potential and therefore performance in WS$_2$ solar cells.

**Section S8. Photoshunting**

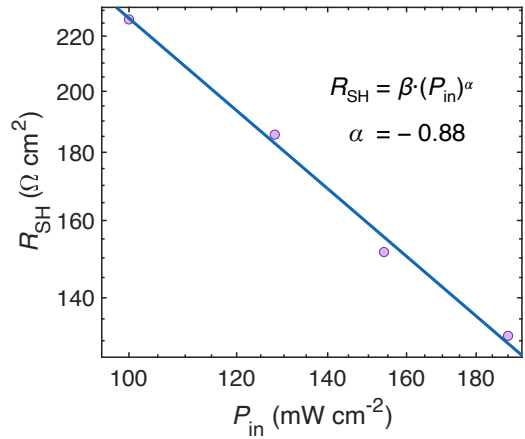

**Figure S8 | Photoshunting.** Shunt resistance ($R_{SH}$) of a typical flexible WSe$_2$ solar cell at various incident power ($P_{in}$) intensities of AM 1.5G illumination. Shunt resistance decreases almost linearly with increasing incident power intensity due to increased minority carrier conductivity across the device under illumination, a phenomenon known as photoshunting.[59,60] Utilizing contacts with greater carrier selectivity or introducing a high built-in potential p-n homojunction could reduce or eliminate the photoshunting observed here. Symbols, measurements; line, power law fit.

**Section S9. Supplementary photocurrent mapping**

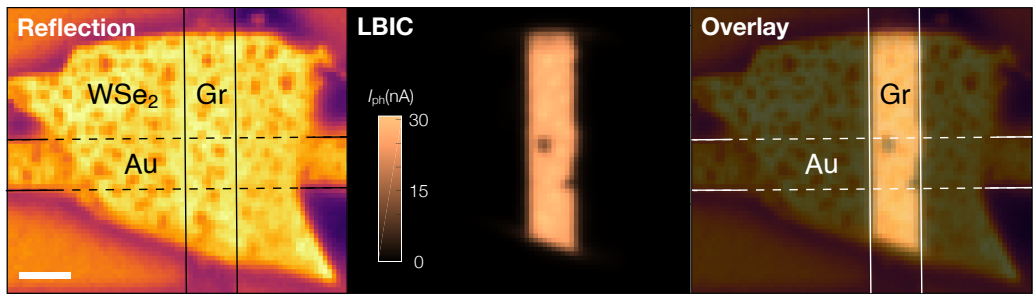

**Figure S9 | Photocurrent mapping.** Spatial maps of reflection, photocurrent, and reflection/photocurrent overlay for a flexible Au–WSe$_2$–Gr solar cell measured at λ = 530 nm. Current generation occurs only in the WSe$_2$ region under the Gr contact, which can be used to accurately define the active area of the solar cell. No photocurrent generation is observed at the Au–WSe$_2$ back diode. Scale bar, 10 µm.

**Section S10. Effects of Gr and MoO$_x$ on optical absorption**

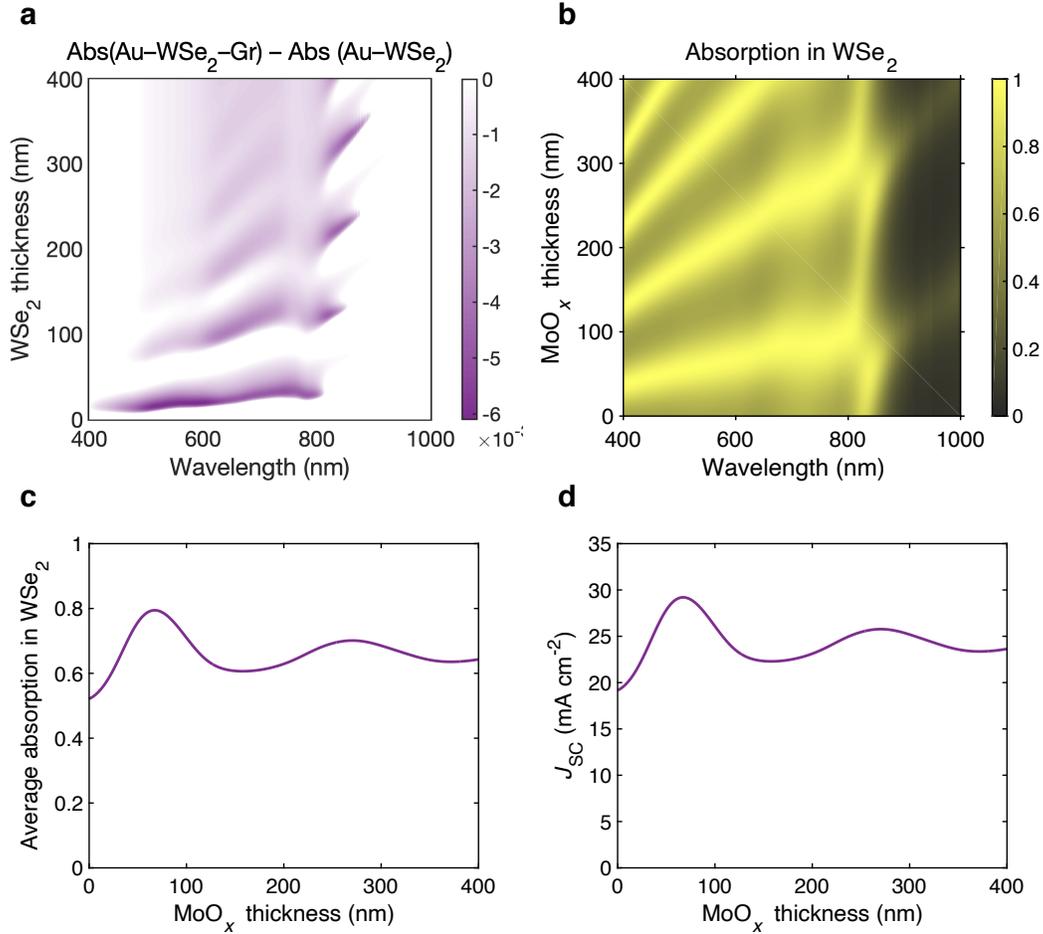

**Figure S10 | Effects of Gr and MoO$_x$ on optical absorption. a,** Absorption reduction by graphene. The difference in total absorption between the Au–WSe$_2$ and Au–WSe$_2$–Gr stacks, showing that adding graphene on top slightly reduces the total absorption in the ~500–800 nm wavelength spectrum for WSe$_2$ thicknesses around 200 nm, in agreement with experimental measurements (Fig. 3c). **b–d,** Anti-reflection coating effects of MoO$_x$. **b,** Absorption spectrum and **c,** average absorption of WSe$_2$ in the Au–WSe$_2$–Gr–MoO$_x$ stack as a function of MoO$_x$ thickness. WSe$_2$ thickness is 209 nm, similar to the device in Fig. 3c. **d,** Maximum short-circuit current density ($J_{SC}$) attainable from WSe$_2$ in the same stack, as a function of MoO$_x$ thickness, calculated by integrating Absorption(λ) × (spectral photon flux of AM1.5G spectrum at one-sun solar intensity) over the wavelength range of λ = 400–1000 nm, assuming unity internal quantum efficiency (IQE). This maximum $J_{SC}$ value is slightly underestimated as absorption at wavelengths below 400 nm and above 1000 nm are not included due to lack of material data. An optimal MoO$_x$ thickness of ~70 nm can increase the average absorption in WSe$_2$ and similarly $J_{SC}$ by ~50%, leading to a remarkable ~80% average absorption and ~30 mA cm$^{-2}$ $J_{SC}$ in WSe$_2$ solar cells.

**Section S11. Benchmarking**

We performed an extensive literature review on flexible and light-weight solar cells with various absorber materials and calculated $P_S$ based on information provided. Some works directly stated $P_S$ or at least PCE and the areal mass density, while others did not explicitly provide information on the weight of their solar cells. In some cases, we calculated the weight based on the substrate and the solar cell layer stack if the substrate weight seemed small (sufficiently thin/low density material). In other cases, we only used the substrate weight for estimating $P_S$ if it appeared to dominate (thicker high-density material substrates) and neglected the weight of other materials. Table S3 lists all the works shown in Fig. 5 indicating how $P_S$ was obtained.

**Table S3 | Literature reports on flexible and light-weight solar cells with notable specific power ($P_S$), along with their power conversion efficiencies (PCE)**. Unmarked $P_S$ is directly taken from the literature reports. *$P_S$ calculated based on the solar cell layer stack including the substrate. **$P_S$ calculated only based on substrate weight or absorber material if free-standing. ***PCE corresponds to performance under AM 0 illumination (space).

| Reference | Technology | PCE (%) | $P_S$ (W g$^{-1}$) |
|---|---|---|---|
| Kaltenbrunner et al.[8] | Perovskite | 12 | 23 |
| Kaltenbrunner et al.[90] | Organic | 4 | 10 |
| Söderström et al.[44] | Amorphous Si (a-Si) | 8.8 | 3.3 |
| Chirilă et al.[42] | CuInGaSe (CIGS) | 18.7 | 3.3 |
| Shiu et al.[91] | InP (III-V) | 10.2 | 2.0 |
| Romeo et al.[41] | CdTe (II-VI) | 11.4 | 2.1 |
| Fatemi et al.[92] | 3-mil Si (c-Si) | 17*** | 0.85 |
| Fatemi et al.[92] | InGaP/GaAs/Ge (III-V) | 26*** | 0.4 |
| Zhao et al.[93] | Single-crystal Si (c-Si) | 24.4 | 0.4 |
| Zhao et al.[93] | Polycrystalline Si (poly-Si) | 19.8 | 0.33 |
| Bremaud et al.[94] | CuInGaSe (CIGS) | 15 | 0.09 |
| Kang et al.[62] | Perovskite | 12.85 | 29.4 |
| Zhang et al.[87] | PbS quantum dot | 9.9 | 15.2 |
| Liu et al.[95] | Perovskite | 11.5 | 5 |
| Lin et al.[4] | Amorphous Si (a-Si) | 7.06 | 0.5** |
| Park et al.[88] | Organic | 10.5 | 11.46 |

| Reference | Technology | PCE (%) | $P_S$ (W g$^{-1}$) |
|---|---|---|---|
| Tavakoli et al.[96] | PbS quantum dot | 7.1 | 12.3 |
| Li et al.[97] | Perovskite | 13 | 3.7** |
| Jinno et al.[98] | Organic | 7.9 | 14* |
| Jia et al.[99] | Perovskite | 18 | 1.3** |
| Lee et al.[100] | Perovskite | 17.03 | 18.5* |
| Cardwell et al.[47] | GaInP/GaAs/GaInAs (III-V) | 29.3*** | 3.8 |
| Xie et al.[101] | Perovskite | 13.32 | 4.16 |
| Sun et al.[5] | Amorphous Si (a-Si) | 5.6 | 1.382 |
| Başol et al.[102] | CuInSe$_2$ (CIS) | 9.3 | 1.133 |
| Rance et al.[103] | CdTe (II-VI) | 14.05 | 0.6** |
| Mahabaduge et al.[104] | CdTe (II-VI) | 16.4 | 0.7** |
| Law et al.[105] | GaInP/GaInAs/Ge (III-V) | 21*** | 2.067 |
| Kim et al.[46] | GaAs (III-V) | 15.2 | 5.9* |
| Gerthoffer et al.[106] | Cu(In,Ga)Se$_2$ (CIGS) | 11.2 | 0.4** |
| Jeong et al.[2] | Single-crystal Si (c-Si) | 13.7 | 5.88** |
| Das et al.[6] | Single-crystal Si (c-Si) | 9 | 1.89 |
| Hwang et al.[107] | Single-crystal Si (c-Si) | 18.4 | 1.3* |
| Jean et al.[108] | Organic | 2.2 | 6.11 |
| Salavei et al.[109] | CdTe (II-VI) | 10 | 1.98 |
| Shahrjerdi et al.[110] | InGaP/(In)GaAs (III-V) | 28.1 | 1.995 |
| Cho et al.[111] | CdS/CdTe (II-VI) | 13.56 | 0.254 |
| Zhao et al.[112] | Organic | 8.7 | 0.4 |
| Li et al.[113] | Perovskite | 14 | 1.96 |
| Romeo et al.[40] | CdTe/CdS (II-VI) | 11 | 2.6* |
| Liu et al.[114] | Organic | 6.62 | 1.71 |
| Chirilă et al.[43] | Cu(In,Ga)Se$_2$ (CIGS) | 20.4 | 3.4* |
| Xu et al.[45] | Amorphous Si (a-Si) | 9.7*** | 1.2 |
| Mavlonov et al.[115] | Cu(In,Ga)Se$_2$ (CIGS) | 9.1 | 0.7* |
| Qu et al.[116] | Organic | 16.1 | 11.5** |
| Koo et al.[117] | Organic | 15.2 | 10.8** |
| Garud et al.[118] | Polycrystalline Si (poly-Si) | 15 | 0.06** |
| El-Atab et al.[119] | Single-crystal Si (c-Si) | 19.1 | 0.24 |

| Reference | Technology | PCE (%) | $P_S$ (W g$^{-1}$) |
|---|---|---|---|
| Augusto et al.[120] | Single-crystal Si (c-Si) | 18.4 | 1.3** |
| Lee et al.[3] | Single-crystal Si (c-Si) | 12.4 | 1* |
| El-Atab et al.[121] | Single-crystal Si (c-Si) | 19 | 0.29 |
| Xue et al.[7] | Single-crystal Si (c-Si) | 12.3 | 9.33* |
| Tanabe et al.[122] | InAs/GaAs quantum dot (III-V) | 10.5 | 0.6** |
| Wu et al.[123] | Dye-Sensitized (DSSC) | 6.69 | 0.3** |
| Akama et al.[20] | WS$_2$ (TMD) | <0.7% | <0.04** |

**Supplementary References:**